\documentclass[fleqn,usenatbib,useAMS]{mnras}
\usepackage{graphicx}
\usepackage{amsmath}
\usepackage{amssymb}
\usepackage{multicol}
\usepackage{verbatim}
\usepackage{pdflscape}
\usepackage{hyperref}
\usepackage[usenames]{color}
\usepackage[T1]{fontenc}
\usepackage{ae,aecompl}
\usepackage{txfonts}
\usepackage{todonotes}
\usepackage{lineno}
\usepackage{ulem}
\graphicspath{{figures/}} 

\newcommand{\dd}{\mathrm{d}}

\newcommand{\redmapper}{{redMaPPer}}
\newcommand{\sdss}{SDSS}  
\newcommand{\planck}{\textit{Planck}}
\newcommand{\chandra}{\textit{Chandra}}
\newcommand{\colossus}{\textsc{Colossus}}

\newcommand{\intr}{\mathrm{intr}}
\newcommand{\gas}{\mathrm{gas}}
\newcommand{\pivot}{\mathrm{pivot}}
\newcommand{\obs}{\mathrm{ob}}
\newcommand{\true}{\mathrm{true}}
\newcommand{\sat}{\mathrm{sat}}

\newcommand{\rmrel}{$\langle \lambda^\true|\mm\rangle$}
\newcommand{\mgmrel}{$\langle M_\gas^\true|\mc\rangle$}
\newcommand{\mm}{M_{\rm 200m}}
\newcommand{\mc}{M_{\rm 500c}}

\newcommand{\msun}{M_\odot}
\newcommand{\hinv}{h^{-1}}
\newcommand{\massunits}{\msun\ \hinv}
\newcommand{\omm}{\Omega_{\rm m}}

\title[Cosmology with a complete cluster sample]
{Improved Cosmological Constraints from SDSS redMaPPer Clusters via X-ray Follow-up of a
Complete Subsample of Systems}

\author[M.~Kirby]{
\parbox{\textwidth}{
\Large
Matthew~Kirby,\thanks{E-mail: matthewkirby@email.arizona.edu}$^{1}$
Eduardo~Rozo,$^{1}$
R.~Glenn~Morris,$^{2,3}$
Steven~W.~Allen,$^{2,4,3}$
Matteo~Costanzi,$^{5,6}$
Tesla~E.~Jeltema,$^{7,8}$
Adam~B.~Mantz,$^{2,4}$
A.~Kathy~Romer,$^{9}$
E.~S.~Rykoff,$^{2}$
and Anja~von~der~Linden$^{10}$
}
\vspace{0.4cm}
\\
\parbox{\textwidth}{
$^{1}$ Department of Physics, University of Arizona, Tuscon, AZ 85721, USA\\
$^{2}$ Kavli Institute for Particle Astrophysics \& Cosmology, P. O. Box 2450, Stanford University, Stanford, CA 94305, USA\\
$^{3}$ SLAC National Accelerator Laboratory, Menlo Park, CA 94025, USA\\
$^{4}$ Department of Physics, Stanford University, 382 Via Pueblo Mall, Stanford, CA 94305, USA\\
$^{5}$ INAF-Osservatorio Astronomico di Trieste, via G. B. Tiepolo 11, I-34143 Trieste, Italy\\
$^{6}$ IFPU - Institute for Fundamental Physics of the Universe, Via Beirut 2, 34014 Trieste, Italy\\
$^{7}$ University of California Santa Cruz, Santa Cruz, CA 95060, USA\\
$^{8}$ Santa Cruz Institute for Particle Physics, Santa Cruz, CA 95064, USA\\
$^{9}$ Astronomy Centre, University of Sussex, Brighton, BN1 9QH\\
$^{10}$ Department of Physics and Astronomy, Stony Brook University, Stony Brook, NY 11794, USA\\
}
}

\pubyear{2019}

\begin{document}
\label{firstpage}
\pagerange{\pageref{firstpage}--\pageref{lastpage}}
\maketitle

\begin{abstract}
We improve upon the cosmological constraints derived from the abundance and weak-lensing data of \redmapper\ clusters detected in the Sloan Digital Sky Survey (\sdss).  Specifically, we derive gas mass data using {\it Chandra} X-ray follow-up of a complete sample of the 30 richest \sdss\ \redmapper\ clusters with $z\in[0.1,0.3]$, and use these additional data to improve upon the original analysis by \citet{Costanzi19_SDSSClusterCosmology}. We simultaneously fit for the parameters of the richness--mass relation, the cluster gas mass--mass relation, and cosmology. By including our X-ray cluster sample in the SDSS cluster cosmology analysis, we measure $\omm = 0.25\pm 0.04$ and $\sigma_8 = 0.85^{+0.06}_{-0.08}$. These constraints represent a 25.5\% and 29.8\% reduction in the size of the 68\% confidence intervals of $\omm$ and $\sigma_8$ respectively, relative to the constraints published in \citet{Costanzi19_SDSSClusterCosmology}. Our cosmological constraints are in agreement with early universe results from \planck.  As a byproduct of our analysis, we also perform an independent calibration of the amplitude of the \mgmrel\ scaling relation.  Our calibration is consistent with and of comparable precision to that of \citet{Mantz16_WtG5}.
\end{abstract}


\setcounter{footnote}{1}

\section{Introduction}
\label{sec:intro}

Galaxy clusters are the largest collapsed objects in the universe and trace the highest peaks in the matter density field. By measuring their abundance as a function of mass and redshift we can place constraints on cosmological parameters, namely the matter density, $\omm$, and the amplitude of matter fluctuations, $\sigma_8$. The $\sigma_8$ constraint from clusters, when combined with a constraint on the amplitude of the matter power spectrum in the early universe --- e.g. from the Cosmic Microwave Background (CMB) --- allows us to place constraints on the growth of structure. Combining these growth of structure constraints with geometry constraints, e.g. from Baryon Acoustic Oscillations (BAO) or Super Nova (SN) data, allows us to use clusters to place strong constraints on the energy density of neutrinos, $\Omega_{\rm \nu}$, the Hubble constant, $h$, and the dark energy equation of state, $w$ \citep{Allen11_ClusterCosmology}.

Cluster abundance analyses have been done for cluster samples selected in optical \citep{Rozo10_SDSSCCMaxBCG, Costanzi19_SDSSClusterCosmology}, X-ray \citep{Vikhlinin09_ChandraClusterCosmology, Mantz10_RASSClusterCosmology, Mantz15_WtG4}, and millimeter wavelengths \citep{Hasselfield13_ActCC, Planck13_ClusterCosmology, Planck16_ClusterCosmology, deHaan16_SPTClusterCosmology, Bocquet18_SPTClusterCosmology}. The cosmological constraints from these experiments have been in general agreement with one another while producing competitive constraints on $\omm$ and $\sigma_8$ dark energy, and modified gravity. Ongoing (e.g. Dark Energy Survey (DES)\footnote{https://www.darkenergysurvey.org}, the Hyper Suprime-Cam\footnote{http://hsc.mtk.nao.ac.jp/ssp/}, the South Pole Telescope\footnote{https://pole.uchicago.edu/}, and the Atacama Cosmology Telescope\footnote{https://act.princeton.edu/}) and future (e.g. Large Synoptic Survey Telescope\footnote{https://www.lsst.org/}, eRosita\footnote{http://www.mpe.mpg.de/eROSITA}, Euclid\footnote{http://sci.esa.int/euclid/}, Simons Observatory\footnote{https://simonsobservatory.org/}, and CMB-S4\footnote{https://cmb-s4.org/}) surveys will increase the number of known galaxy clusters by an order of magnitude or more.

The principal difficulty for cluster cosmology is that cluster mass is not easily observable. Weak gravitational lensing is the current gold standard for absolute mass calibration. While the signal to noise of individual cluster measurements is typically below one for low mass clusters, we can increase the signal-to-noise ratio through cluster stacking. We then use the resulting stacked weak lensing masses to calibrate mean-mass scaling relations for other cluster observables. The mass proxies most commonly used are richness for optically selected clusters \citep[e.g.][]{Simet17_SDSSMassCalibration, Melchior17_DESSVWeakLensingMasses, Murata18_SDSSWeakLensingMasses,  McClintock19_DESY1MassCalibration}, X-ray gas mass, temperature, and luminosity  for clusters selected in the X-rays \citep[e.g.][]{vonderLinden14_WtGWeakLensing, Applegate14_WtGWeakLensing, Mantz16_WtG5, Dietrich19_SPTSelectedScaling}, and the thermal Sunyaev-Zeldovich (tSZ) signal for millimeter wavelength surveys \citep[e.g.][]{Planck16_ClusterCosmology, deHaan16_SPTClusterCosmology, Bocquet18_SPTClusterCosmology, Dietrich19_SPTSelectedScaling}. Each of these proxies have their own strong points: richness is observationally cheap, X-ray gas mass and temperature are expected to be tightly correlated with halo mass, and the tSZ signal is redshift-independent, allowing for the creation of approximately mass-limited samples.

While weak gravitational lensing of stacks of clusters is effective at calibrating the mean of the observable--mass relation, the uncertainty of the lensing mass for a single cluster is large. Since this uncertainty is larger than or comparable to the intrinsic scatter of the observable, weak lensing cannot currently be used to measure the scatter. Without a tight constraint on the scatter, the cosmological posteriors of a cluster cosmology experiment are broadened. However, by adopting an unbiased, low-scatter mass proxy for a complete sample of clusters, one can recover this lost information, thereby improving the final cosmological constraints. Here, we perform a simultaneous analysis of the abundance and weak lensing data from optically selected clusters in SDSS alongside X-ray gas masses derived from \chandra\ observations for the richest 30 clusters.

As this paper was being completed, currently unpublished work from the Dark Energy Survey collaboration demonstrated that the impact of selection effects in \citet{Simet17_SDSSMassCalibration} (the source of our weak lensing mass calibration) was under-estimated (Costanzi et al, in preparation).  Adopting these new corrections increases our systematic error budget, but should not otherwise impact our conclusions, particularly with regards to the utility of multi-wavelength cluster data for calibration of the scatter in the richness--mass relation, and the corresponding improvement in the cosmological posteriors. 

This paper is organized as followed. In Section~\ref{sec:data}, we describe the various data sets we employ in our analysis, as well as the data products of the analysis of \citet{Costanzi19_SDSSClusterCosmology}, upon which our analysis builds. Section~\ref{sec:xray-profiles} collects the observational systematics that are accounted for in our work. We describe our theoretical model in Section~\ref{sec:lkhd}. In Section~\ref{sec:results}, we present our results. Unless otherwise noted, all of our masses are spherical overdensity (SO) masses defined with respect to an average density of $200$ times the mean matter density of the universe, $\mm$. Cluster gas mass is defined within the radius $R_{\rm 500c}$, the radius within which the average density of the cluster is $500$ times the critical density of the universe. We define our reference cosmological model to be a flat $\Lambda$CDM cosmological model with a Hubble parameter $h=0.70$ and local mean matter density $\omm = 0.30$.

\section{Data Sets}
\label{sec:data}

\begin{table*}
    \centering\footnotesize
    \caption{The 30 richest clusters in the SDSS \redmapper\ cluster catalog. We note that center-corrected richness for one of our clusters, Abell~1095, dropped below the richness of the 30th richest cluster. Asterisks indicate the 11 clusters for which we obtained new \chandra\ imaging. 
(1) \redmapper\ ID.
(2) Cluster name. 
(3) Right ascension of the \redmapper\ center (J2000).
(4) Declination of the \redmapper\ center (J2000).
(5) Cluster gas mass within $R_{500c}$ in our reference cosmology in units of $10^{14} \msun$. 
(6) Richness at the \redmapper\ center for each of our clusters. 
(7) ``Center-Corrected'' richness at the X-ray center for each of our miscentered clusters. 
(8) The \redmapper\ photometric redshift of the cluster. 
(9) Classification from visual inspection of the X-ray images.}
\begin{tabular}{rlccccccl}
	\hline
	ID & Cluster Name & R.A. & Decl. & $M^{\rm ref}_{\gas}$ & $\lambda$ & $\lambda^{\rm CC}$ & $z$ & Classification  \\
	 & & (J2000) & (J2000) & $\left(10^{14}\,\msun\right)$ & & & \\
    	\hline \vspace{-3mm}\\
	2 & Abell 2219 & 16h40m19.812s & +46d42m41.51s & $2.55 \pm 0.26$ & $202.6$ & \--- & $0.233$ & Well-Centered, Unimodal \vspace{0.5mm} \\
	6 & Abell 1763 & 13h35m20.093s & +41d00m04.13s & $1.88 \pm 0.21$ & $191.0$ & \--- & $0.232$ & Well-Centered, Unimodal \vspace{0.5mm} \\
	14 & Abell 2261 & 17h22m27.182s & +32d07m57.25s & $1.28 \pm 0.14$ & $185.8$ & \--- & $0.229$ & Well-Centered, Unimodal \vspace{0.5mm} \\
	5 & Abell 750 & 09h09m12.179s & +10d58m24.94s & $0.83 \pm 0.12$ & $176.2$ & \--- & $0.170$ & Well-Centered, Unimodal \vspace{0.5mm} \\
	23 & Abell 781 & 09h20m25.790s & +30d29m38.70s & $1.11 \pm 0.11$ & $173.9$ & \--- & $0.299$ & Well-Centered, Unimodal \vspace{0.5mm} \\
	3 & Abell 1689 & 13h11m29.510s & -01d20m28.02s & $1.29 \pm 0.12$ & $165.3$ & \--- & $0.182$ & Well-Centered, Unimodal \vspace{0.5mm} \\
	34 & Abell 1319$^*$ & 11h36m13.045s & +40d02m35.78s & $0.70 \pm 0.08$ & $158.3$ & \--- & $0.296$ & Recent Merger \vspace{0.5mm} \\
	16 & Abell 1703 & 13h15m05.236s & +51d49m02.80s & $1.12 \pm 0.11$ & $155.2$ & \--- & $0.284$ & Well-Centered, Unimodal \vspace{0.5mm} \\
	15 & Abell 773 & 09h17m53.420s & +51d43m37.54s & $1.01 \pm 0.10$ & $154.6$ & \--- & $0.226$ & Well-Centered, Unimodal \vspace{0.5mm} \\
	40 & Abell 1758 & 13h32m38.411s & +50d33m36.00s & $1.29 \pm 0.13$ & $152.8$ & \--- & $0.290$ & Recent Merger \vspace{0.5mm} \\
	19 & Abell 2111 & 15h39m40.493s & +34d25m27.29s & $0.95 \pm 0.11$ & $152.2$ & \--- & $0.233$ & Well-Centered, Unimodal \vspace{0.5mm} \\
	58 & Abell 1622$^*$ & 12h49m30.924s & +49d49m02.32s & $0.51 \pm 0.15$ & $147.5$ & \--- & $0.284$ & Recent Merger \vspace{0.5mm} \\
	28 & Abell 2390 & 21h53m36.831s & +17d41m43.73s & $2.26 \pm 0.33$ & $146.0$ & \--- & $0.251$ & Well-Centered, Unimodal \vspace{0.5mm} \\
	31 & Abell 697 & 08h42m57.556s & +36d21m59.26s & $1.93 \pm 0.22$ & $141.2$ & \--- & $0.297$ & Well-Centered, Unimodal \vspace{0.5mm} \\
	44 & SDSSJ 1233+1511$^*$ & 12h34m16.148s & +15d15m08.41s & $0.64 \pm 0.14$ & $138.8$ & $124.3$ & $0.281$ & Grossly Miscentered \vspace{0.5mm} \\
	43 & Abell 1835 & 14h01m00.528s & +02d51m49.71s & $1.56 \pm 0.15$ & $135.6$ & $142.6$ & $0.257$ & Grossly Miscentered \vspace{0.5mm} \\
	11 & Abell 655 & 08h25m29.062s & +47d08m00.86s & $0.36 \pm 0.05$ & $134.6$ & \--- & $0.127$ & Well-Centered, Unimodal \vspace{0.5mm} \\
	48 & RMJ 1603.3+0316.7$^*$ & 16h03m18.986s & +03d16m44.57s & $0.46 \pm 0.06$ & $133.8$ & \--- & $0.237$ & Well-Centered, Unimodal \vspace{0.5mm} \\
	59 & SDSSJ 1201+2306 & 12h01m12.199s & +23d05m57.30s & $1.49 \pm 0.12$ & $133.3$ & \--- & $0.269$ & Recent Merger \vspace{0.5mm} \\
	49 & RMJ 2307.1+1632.8$^*$ & 23h07m07.499s & +16d32m46.06s & $1.00 \pm 0.15$ & $132.2$ & \--- & $0.248$ & Recent Merger \vspace{0.5mm} \\
	57 & SDSSJ 0104+0003 & 01h04m55.371s & +00d03m36.34s & $0.71 \pm 0.14$ & $129.7$ & \--- & $0.277$ & Well-Centered, Unimodal \vspace{0.5mm} \\
	46 & Abell 1095$^*$ & 10h47m29.011s & +15d14m02.08s & $0.71 \pm 0.19$ & $129.6$ & $97.2$ & $0.209$ & Grossly Miscentered \vspace{0.5mm} \\
	13 & Abell 98 & 00h46m29.328s & +20d28m04.85s & $0.25 \pm 0.03$ & $129.2$ & \--- & $0.104$ & Recent Merger \vspace{0.5mm} \\
	24 & RMJ 1334.1+2014.9$^*$ & 13h34m08.681s & +20d14m53.00s & $0.63 \pm 0.11$ & $126.4$ & \--- & $0.171$ & Well-Centered, Unimodal \vspace{0.5mm} \\
	107 & RMJ 1054.3+1439.1$^*$ & 10h54m17.543s & +14d39m04.21s & $0.97 \pm 0.19$ & $125.4$ & \--- & $0.298$ & Well-Centered, Unimodal \vspace{0.5mm} \\
	75 & Abell 959$^*$ & 10h17m34.326s & +59d33m39.82s & $0.69 \pm 0.13$ & $125.2$ & \--- & $0.289$ & Well-Centered, Unimodal \vspace{0.5mm} \\
	69 & Abell 1682 & 13h06m45.691s & +46d33m30.77s & $0.74 \pm 0.09$ & $124.8$ & $125.9$ & $0.229$ & Grossly Miscentered \vspace{0.5mm} \\
	22 & Abell 586 & 07h32m20.315s & +31d41m20.70s & $0.64 \pm 0.07$ & $122.4$ & $145.9$ & $0.180$ & Grossly Miscentered \vspace{0.5mm} \\
	66 & RMJ 2118.8+0033.6$^*$ & 21h18m49.070s & +00d33m37.23s & $0.40 \pm 0.07$ & $121.0$ & \--- & $0.265$ & Recent Merger \vspace{0.5mm} \\
	42 & RMJ 1643.4+1322.6$^*$ & 16h43m25.376s & +13d22m35.89s & $0.27 \pm 0.06$ & $116.9$ & \--- & $0.187$ & Well-Centered, Unimodal \vspace{0.5mm} \\
	\vspace{0.5mm} \\    		\hline \vspace{-3mm}\\
\end{tabular}
    \label{tab:data:cluster-summary}
\end{table*}

\subsection{The \sdss\ \redmapper\ catalog}
\label{sec:data:redmapper}

The \redmapper\ algorithm is a red-sequence cluster finder designed and optimized for large photometric surveys \citep{Rykoff14_RM1}. The algorithm iteratively finds galaxy clusters by identifying overdensities of red-sequence galaxies and computes the probability that each galaxy is a cluster member. For each cluster, \redmapper\ computes the richness ($\lambda$) by taking the probability-weighted sum of the galaxies in the cluster. The \redmapper\ richness correlates well with other mass proxies such as X-ray luminosity and SZ signal \citep{Rozo14_RM2}.

\redmapper\ was run on $\sim$10,000 square degrees of the \sdss\ Data Release 8\footnote{https://www.sdss3.org/dr8/} (DR8) data and found  $\sim$27,000 galaxy clusters between $0.05 \le z \le 0.6$. We use version 5.1 of the \sdss\ DR8 \redmapper\ catalog to construct a complete, volume-limited sample of clusters by including only the clusters between $0.1 \le z \le 0.3$. This leaves us with $\sim$7000 clusters. We selected the 30 richest clusters for our multi-wavelength analysis.

\subsection{Chandra X-ray gas mass}
\label{sec:data:mgas}

\begin{figure}
    \centering
    \includegraphics[width=\linewidth]{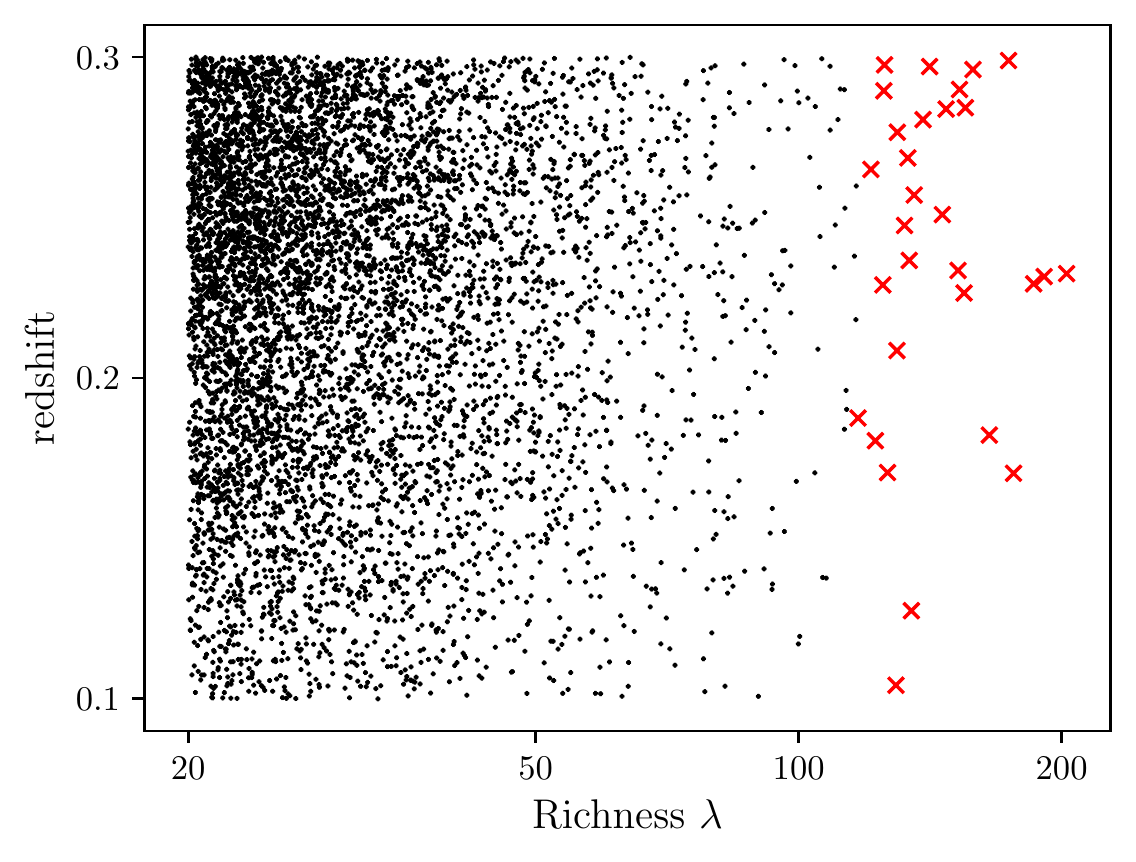}
    \caption{The SDSS Data Release 8 \redmapper\ cluster catalog used in the SDSS cluster cosmology analysis \citep{Costanzi19_SDSSClusterCosmology}. In red, we show the richest 30 clusters for which we have X-ray gas masses.}
    \label{fig:all-sdss-clusters}
\end{figure}

\begin{figure}
    \centering
    \includegraphics[width=\linewidth]{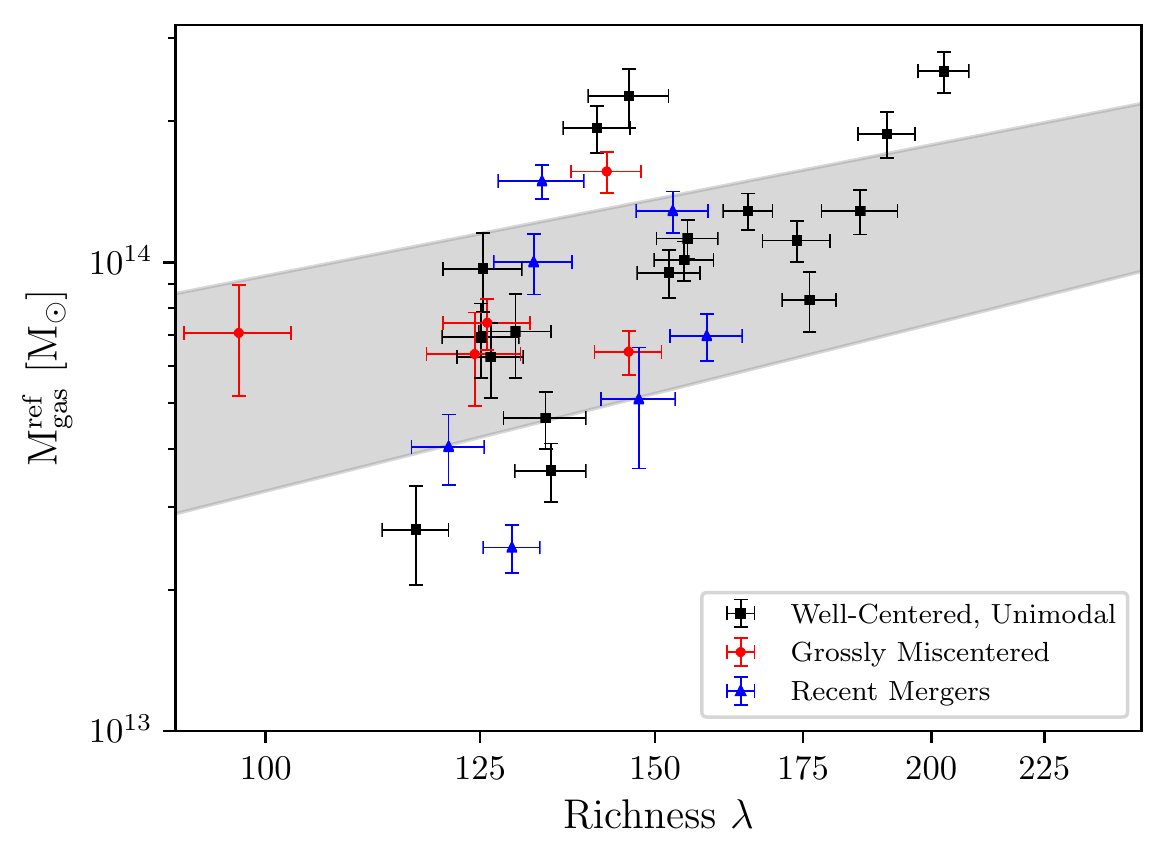}
    \caption{X-ray gas mass and richness for each of the 30 richest clusters in \sdss\ \redmapper\ catalog. The gray band is the 68\% confidence region derived from our best fit model. We have classified each cluster by the features of their X-ray images and show each as a different color and symbol. In Section~\ref{sec:xray-profiles} we describe the different methods that we use to compute the richness and gas mass for each classification.}
    \label{fig:data}
\end{figure}

We obtained \chandra\ imaging for 11 of the 30 richest \sdss\ \redmapper\ clusters with redshifts $z~\in~[0.1,0.3]$, comprising a total of 1.3~Msec over 63 exposures. These new observations, alongside \chandra\ archival data, enabled us to have deep X-ray observations for each of the 30 richest clusters. Figure~\ref{fig:all-sdss-clusters} shows the full \sdss\ \redmapper\ cluster catalog after applying the $z~\in~[0.1, 0.3]$ and $\lambda^\obs~\in~[20, 140)$ selection cuts used in \citet{Costanzi19_SDSSClusterCosmology}, and the 30 richest clusters that comprise our complete X-ray follow-up subsample.  Figure~\ref{fig:data} shows the relation between X-ray and optical cluster observables and the corresponding scaling relation.  The data for our complete subsample of clusters is summarized in Table~\ref{tab:data:cluster-summary}.

We reduced the data and computed the gas mass for each cluster assuming our reference cosmology following the method detailed in \citet{Mantz16_WtG5}, henceforth, M16. In Section~\ref{sec:lkhd:obsvar}, we describe how we convert our predictions of the gas mass from an arbitrary cosmology to our reference cosmology.  We note that 13 of our 30 clusters had either unusual X-ray profiles or were clearly miscentered by \redmapper. In Section~\ref{sec:xray-profiles} we describe how we handled each of these clusters, which sometimes included re-estimating the richness of a cluster at the X-ray center.  We found no significant differences in our results whether we used the richness values at the original cluster centers or the center-corrected richness estimates.

\subsection{\sdss\ cluster cosmology}
\label{sec:data:chains}

\citet{Costanzi19_SDSSClusterCosmology}, henceforth C19, performed a blinded analysis of cluster abundances and weak lensing data from the \sdss\ DR8 \redmapper\ cluster catalog.  Specifically, C19 relied on the mass calibration from \citet{Simet17_SDSSMassCalibration} to jointly constrain cosmology and the richness--mass relation. The mass calibration of \citet{Simet17_SDSSMassCalibration} was performed via a stacked weak lensing analysis in four bins of observed richness. This analysis constrained the mean of the richness--mass relation while accounting for photometric redshift errors, cluster miscentering, shear bias, blending, triaxiality, and projection effects.  We will refer to the results of the cluster weak-lensing and number counts analysis in C19 as the ``SDSS Clusters'' constraints.  We note that our current analysis incorporates the abundance data of our richest clusters, so the abundance data of our systems is ``double counted'' when combining with the C19 analysis.  We have explicitly verified that this amount of extra information is negligible by rerunning the C19 analysis without the abundance information of the richest bin (which extends to $\lambda\geq 69.3$), and verifying that our posteriors are not changed in any significant way.

In addition to the SDSS Clusters constraints detailed above,  C19 also derived cosmological constraints after adding Baryon Acoustic Oscillation (BAO) data to constrain the Hubble constant. The BAO data comes from the 6dF galaxy survey \citep{Beutler11_6dF}, the \sdss\ Data Release 7 Main Galaxy sample \citep{Ross15_SDSSGalaxyClustering} and the Baryon Oscillation Spectroscopic Survey \citep{Alam17_BOSS}. We will refer to the results of the analysis of cluster weak-lensing and number counts with BAO data in C19 as ``Clusters + BAO''.

Finally, C19 derived cosmological constraints including the SDSS clusters weak lensing and number counts data, the BAO data detailed above, and CMB data. The CMB data is from \planck's 2015 data release (DR15) \citep{Planck15_Cosmology} and includes low-$\ell$ polarization data. We will refer to the results of the analysis of cluster weak-lensing and number counts with BAO and CMB data in C19 as ``Clusters + BAO + Planck15''.

We note that our comparison to \planck\ relies on the value of $\sigma_8$, a parameter that needs to be extrapolated from early universe CMB measurements. While $\sigma_8$ cannot be directly measured from CMB primary fluctuations, structure formation models allow us to extrapolate the measurement of the amplitude of temperature fluctuations in the CMB to the amplitude of the linear matter density field today.  Any tension in the values of $\omm$ and $\sigma_8$ between our results and those derived from the CMB could point to either a failure of the standard cosmological model or unmodeled systematics. We use the TT+lowP constraints from \planck's 2015 data release \citep{Planck15_Cosmology}, to make this comparison.  We rely on \planck\ 2015 rather than \planck\ 2018 because the latter has not yet been incorporated into the \textsc{CosmoSIS} framework utilized in our analysis \citep{Zuntz15_Cosmosis}.


\section{Observational Systematics}
\label{sec:xray-profiles}

The X-ray and optical data we use are each subject to various observational systematics that need to be accounted for.  Below, we start by discussing systematics in the optical data, and then move on to discuss systematics in the X-ray data, each of which may impact the recovered richness and gas mass of a given cluster.

During cluster-finding, \redmapper\ assigns a central galaxy to each cluster. This choice is iterative and probabilistic and there is no guarantee that it made the correct choice. As the name suggests, \redmapper\ relies heavily on galaxy color.  Therefore, any effects (instrumental or astrophysical) that impact the color of the cluster members may result in an incorrect choice for the central galaxy. Phenomena that can impact central galaxy assignment include problems in survey photometry (e.g. insufficient masking near a bright source), Active Galactic Nuclei (AGN) emission, unexpected bursts of star formation, and recent cluster mergers where both progenitors had similar central galaxies. Assigning a different central galaxy will result in a different richness. However, X-ray imaging of the intra-cluster gas allows us to more reliably determine the correct central galaxy. We found that the X-ray profiles for 17 of our 30 clusters were unimodal and peaked on the galaxy that \redmapper\ flagged as the central galaxy. We classify these clusters as well-centered unimodal and we do not adjust their richness or gas mass.

We also found 5 clusters that appear unimodal in their X-ray images, but for which the \redmapper\ centers were over 150~kpc away from the X-ray center in our reference cosmology. For each of these clusters, we re-estimated the cluster richness at the X-ray center. We performed our analysis first using the original \redmapper\ richnesses and then again using the ``center-corrected'' richnesses ($\lambda^{\rm CC}$) and found no significant differences between our results. We present both the original \redmapper\ richness of each cluster and our center-corrected richnesses in Table~\ref{tab:data:cluster-summary}. In the following, we present our results using the center-corrected richness.

It is also important to understand why \redmapper\ got each of the five cluster centers above wrong. One cluster (SDSSJ~1233+1511) has a massive, unmasked foreground galaxy affecting galaxy colors that compromises the photometry in about one quarter of the cluster area. Visually, the density of cluster members behind the foreground galaxy is comparable to the density in the unaffected areas. After rerunning \redmapper\ at the X-ray center, we multiplied the richness by 4/3 to correct for the richness contribution of the compromised survey area. Two of the miscentered clusters (Abell~1835 and Abell~1095) had X-ray signals that were centered on galaxies with sufficiently strong emission lines that they pushed the color of the central galaxy away from the red-sequence. One of these two galaxies hosts an Active Galactic Nuclei (AGN), while the other is a starbursting galaxy. For Abell~1095, after reestimating the richness at the X-ray center, its richness falls below that of the 30th richest cluster. Consequently, we omit this cluster from our analysis. 

SDSS did not measure spectra for the central galaxies of the last two miscentered clusters. For one case (Abell~1682), visual inspection reveals two obvious central galaxy candidates. These two galaxies were also identified by \redmapper\ to be the top two most likely central candidates. However, the X-ray emission is centered on the galaxy \redmapper\ labeled as the second best central candidate. We measure the center-corrected richness at the X-ray center. Visual inspection of the optical image for the second cluster, Abell~586, reveals that the cluster has an obvious central galaxy which was not identified by \redmapper. The true central is much bigger and brighter than any other object in the field, however, \redmapper\ did not flag it as a top 5 central galaxy candidate. Upon inspection in the SDSS skyserver, we identified a clear problem in the photometry of the obvious central galaxy. This is evidenced by an SDSS photometric redshift of $\approx0.1$ compared to $\approx0.2$ for the other member galaxies. However, we are unable to establish whether the color has been compromised by astrophysical effects such as star formation or AGN emission or by bad photometry. We again measured the center-corrected richness at the X-ray center.

We also found that seven of our clusters have two distinct X-ray peaks, which we interpret as evidence of a recent merger. For five of these seven, the X-ray peaks visually overlapped. Since we have defined $M_{\rm gas}$ to be the total gas mass within $R_{500c}$, any features of the gas density profile at radii less than $R_{500c}$ do not have a sizable impact on our measurement. Therefore, we chose to make the gas mass measurement centered at the point that maximizes the symmetry of the profile. The X-ray peaks for the other two clusters (Abell~98 and Abell~1622) are well-separated (>500~kpc in our reference cosmology). For these two clusters, we estimated the gas mass at each peak independently. After confirming that the smaller of the two peaks fell within $R_{500c}$ of the larger peak, we add the two measurements together to measure the total cluster gas mass. For all seven these clusters, we re-estimated the richness at each of the X-ray peaks. We found that for all seven clusters these measurements differed by less than 2\% relative to both each other and the richness at the original center. For each of these clusters, we use the original \redmapper\ richness.

\section{Likelihood Model}
\label{sec:lkhd}

We adopt a multi-observable Poisson likelihood
\begin{align}\label{eq:likelihood}\begin{split}
\ln \mathcal{L}_{\rm Chandra} = 
& \sum_{i} \ln \left. \frac{dN (\lambda^\obs, M_\gas^\obs | \boldsymbol{p})}{d\lambda^\obs \,dM_\gas^\obs}\right|_{\lambda^\obs_i, \,M^\obs_{\gas, i}} \\
& - \int_{\lambda_{\rm N}}^\infty d\lambda^\obs \int_0^\infty dM_\gas^\obs \frac{dN (\lambda^\obs, M_\gas^\obs|\boldsymbol{p})}{d\lambda^\obs \,dM_\gas^\obs} \\
& + \mathrm{const.}
\end{split}\end{align}
where the sum runs over the clusters in our sample, $\boldsymbol{p}$ is the vector of our model parameters, and $\lambda_{\rm N}$ is the richness of the least rich cluster in our sample. The first term represents the likelihood of observing a cluster with its observed properties given our model. The second term represents the total number of clusters in the sample given our cluster selection function. In the following sections, we build up $dN/d\lambda^\obs \,dM_\gas^\obs$ beginning from the scaling relations for the true cluster observables. We provide a complete derivation and justification of our likelihood model in Appendix~\ref{app:Lderivation}. In Appendix~\ref{app:validation}, we describe the method that we use to validate our likelihood model using mock cluster catalogs.

\subsection{The scaling relations}
\label{sec:lkhd:scalingrels}

We parameterize the richness--mass relation as in C19:
\begin{equation}
    \label{eq:richness-mass-relation}
    \langle \lambda^{\sat} | \mm, \boldsymbol{p} \rangle = 
    	\left( \frac{\mm - M_{\rm min}}{M_1 - M_{\rm min}} \right)^{\alpha_\lambda}
\end{equation}
Where the expected true richness of a cluster is $\langle \lambda^\true | \mm, \boldsymbol{p} \rangle = 1 + \langle \lambda^\sat | \mm, \boldsymbol{p} \rangle$. The first term describes the number of expected central galaxies in the cluster, which is always one at the halo masses relevant for this work. The second term is the expected number of satellite galaxies in the cluster. $M_\mathrm{min}$ is the minimum mass for a halo to form a central galaxy and $M_1$ is the mass at which a halo obtains its first satellite galaxy. The scatter about the richness--mass relation is modeled as the sum in quadrature of a Poisson term and an intrinsic variance term, 
\begin{equation}
    \sigma^2_\lambda = \langle \lambda^\sat | \mm, \boldsymbol{p} \rangle 
    + \langle \lambda^\sat | \mm, \boldsymbol{p} \rangle^2 \sigma_{\lambda,\intr}^2
\end{equation}

The gas mass--mass scaling relation is modeled as a power law in $\mc$ with Gaussian intrinsic scatter,
\begin{align}
     \label{eq:mgas-m}
     & \langle M_\gas^\true | \mc, \boldsymbol{p} \rangle = 
     	A_\gas \left( \frac{\mc}{M_\pivot\,h_{70}} \right)^{B_\gas} \\
     & \sigma^2_\gas =\langle M_\gas | \mc, \boldsymbol{p} \rangle^2 \sigma_{\gas,\intr}^2 .
\end{align} 
Where $M_\pivot = 7\times 10^{14}\ \massunits$ ($=10^{15}\ \msun$ in our reference cosmology). To convert between spherical overdensity mass definitions, we use the universal concentration--mass relation developed by \citet{Diemer15_CMRelation} and implemented in the open-source python package \colossus\ \citep{Diemer18_Colossus}.

Finally, we allow for correlation between the two observables, such that $P(\lambda^\true, M_\gas^\true|M, \boldsymbol{p})$ is Gaussian with a covariance matrix
\begin{equation}
    C =\begin{bmatrix}
        \sigma^2_\lambda & r\,\sigma_\lambda\,\sigma_\gas \\
        r\,\sigma_\lambda\,\sigma_\gas & \sigma^2_\gas
    \end{bmatrix}
\label{eq:cov_in}
\end{equation}

\subsection{Measurement uncertainty}
\label{sec:lkhd:obsvar}

We model the observed properties of galaxy clusters as random variables described by an expectation value with an associated scatter. This scatter is a combination of the true physical scatter and uncertainty introduced by our observations. 

We model $P(M_\gas^\obs|M_\gas^\true)$ as Gaussian, centered on $M_\gas^\true$, with a variance given by the measurement uncertainty. We analytically convolve this distribution with the distribution $P(\lambda^\true,M_\gas^\true|M,\boldsymbol{p})$ described above. Because we define gas mass in terms of $R_{500c}$, this measurement necessarily includes an assumption about cosmology. When running our chains, we rescale the predicted $M_\gas$ to the value that would be inferred in our fiducial cosmology before comparing to $M_\gas^\obs$.  The conversions are done following \citet{Mantz16_WtG5}.

\cite{Costanzi19_Projections} quantifies observational effects on the \redmapper\ richness by combining data analysis and simulated halo catalogs. Incorporating the effects of correlated large scale structure, uncertainty in the background subtraction during cluster finding, and projection effects due to the detection and percolation of nearby clusters, they report a model for the distribution of observed richness at a fixed true richness and redshift, $P(\lambda^\obs|\lambda^\true, z)$. We evaluate this distribution at the median redshift of the \redmapper\ cluster catalog, $z=0.22$. We have opted to ignore this redshift evolution since the redshift range of our survey is narrow and the evolution is weak. In Appendix~\ref{app:validation}, we have tested this approximation using mock cluster catalogs and find that it does not bias our results. We convolve this distribution with $P(\lambda^\true, M_\gas^\obs|M)$ to arrive at the distribution for the cluster observables at fixed mass. 
\begin{equation}
    \label{eq:p_obs_m}
    P(\lambda^\obs, M_\gas^\obs|M, \boldsymbol{p}) = 
    	\int_0^\infty \dd\lambda^\true P(\lambda^\obs|\lambda^\true)\; P(\lambda^\true, M_\gas^\obs|M, \boldsymbol{p})
\end{equation}

\subsection{Halo mass function}
\label{sec:lkhd:massfunc}

We adopt the Tinker mass function \citep{Tinker08_MassFunction} in this work. \cite{Tinker08_MassFunction} reports that their mass function is calibrated to $\lesssim 5\%$ uncertainty over the mass range $10^{11}\ \massunits < M < 10^{15}\ \massunits$. To model this uncertainty, we follow the prescription used in C19 and introduce two nuisance parameters $s$ and $q$ to model the true halo mass function via
\begin{equation}
    \label{eq:true_mf}
    n(M, \boldsymbol{p}) = n(M)^{\mathrm{Tinker}}\, \left( s \log_{10} \left(\frac{\mm}{M^*} \right) + q \right) .
\end{equation}
We use a pivot mass $\log_{10} M^* = 13.8 \,\massunits$. For the \sdss\ cosmology chains, C19 derived a bivariate Gaussian prior on $s$ and $q$ described with means values of $\bar{s} = 0.037$ and $\bar{q} = 1.008$ and a covariance matrix,
\begin{equation}
    C(s,q) = 
    \begin{bmatrix}
        0.00019 & 0.00024 \\
        0.00024 & 0.00038
    \end{bmatrix}.
\end{equation}

C19 calibrated the mean and covariance of the two nuisance parameters using the dark matter only simulations done for the Aemulus project \citep{Derose19_Aemulus}. These 40 simulations span a range of cosmologies and include particles with redshifts $0.0 < z < 1.0$. The \texttt{ROCKSTAR} algorithm \citep{Behroozi13_Rockstar} was used to find the halos in the simulations which were then used to calibrate the priors. See C19 for the full calibration method.

The Tinker mass function was calibrated on dark matter n-body simulations for many cosmologies with massless neutrinos. \citet{Costanzi13_Neutrinos} showed that the Tinker mass function remains a valid approximation for a universe with massive neutrinos if, while computing the mass function, one both: (1) neglects the contribution of neutrinos to the energy density while computing the relation between mass and scale and (2) considers only the cold dark matter and baryon components of the power spectrum. We make these approximations to incorporate massive neutrinos in our analysis.

Finally, we convolve equations~\ref{eq:p_obs_m} and~\ref{eq:true_mf} to calculate the cluster abundance function, $\dd N/\dd\lambda^\obs \dd M_\gas^\obs$, the expected number of cluster per observed richness and gas mass.
\begin{align}
    \label{eq:p_obs}
    \frac{\dd N (\lambda^\obs, M_\gas^\obs | \boldsymbol{p})}{\dd\lambda^\obs \dd M_\gas^\obs} = 
    	V_{\rm s}(\boldsymbol{p}) \int_0^\infty \dd M\;n(M, \boldsymbol{p})\, 
	P(\lambda^\obs, M_\gas^\obs | M, \boldsymbol{p})
\end{align}
where $V_{\rm s}(\boldsymbol{p})$ is the comoving survey volume of the \sdss\ DR8 \redmapper\ catalog in a cosmology defined by $\boldsymbol{p}$.

We note that in our analysis we do not account for the redshift evolution of the mass function. In each cosmology, we compute the mass function at the pivot redshift of our cluster sample, $z=0.22$. In Appendix~\ref{app:validation}, we describe the technique that we use to validate this approximation using mock cluster catalogs built assuming a fiducial model with the full redshift evolution. We find that this approximation has not introduced any biases into our results.

\subsection{Model parameters}
\label{sec:lkhd:params}

In order to derive cosmological constraints from the \sdss\ sample, C19 performed a Markov Chain Monte Carlo  (MCMC) analysis combining cluster abundance data and weak lensing mass estimates. The chain sampled the cosmological parameters ($\omm$, $h_0$, $\Omega_{\rm b}\,h^2$, $\log(10^{10} A_s)$, $n_{\rm s}$, $\Omega_{\rm \nu}\,h^2$, $\sigma_8$), mass function nuisance parameters ($s$, $q$), and the richness--mass relation parameters ($\log M_1$, $\log M_\mathrm{min}$, $\alpha$, $\sigma_{\lambda,\intr}$).  We extended C19's chains into the parameter space describing the gas mass--mass relation using X-ray priors as described below.

For each of the X-ray parameters, we define a prior from which we draw a value for each link in C19's MCMC chains. For the amplitude and slope of the gas mass--mass relation, we define conservative priors that are flat over the ranges $A_\gas/M_{\pivot} \in [0.5\Omega_{\rm b}/\Omega_{\rm m}, \Omega_{\rm b}/\Omega_{\rm m}]$ and $B_\gas \in [1.0, 1.4]$. These ranges represent conservative bounds on the scaling relation parameters that span measurements on X-ray \citep{Mantz16_WtG5} and SZ \citep{Chiu18_SPTBaryonContent, Bulbul19_SPTXMM, Dietrich19_SPTSelectedScaling} selected cluster samples and on high mass clusters in hydrodynamic simulations \citep{Battaglia13_Hydro, LeBrun17_Hydro, Truong18_Hydro}.

We select the prior on $\sigma_{\gas,\intr}$ to be uniform over $(0.00, 0.11]$. This upper limit was chosen as it lies at the upper edge of the confidence interval when the scatter was measured using hydrostatic mass estimates of relaxed clusters \citep{Mantz16_WtG3}. We caution that by construction, this scatter does not explicitly test scatter that arises in the case of merging clusters, as discussed in section~\ref{sec:xray-profiles}.

To define the prior on the correlation coefficient between the intrinsic richness and intrinsic gas mass we consider the boundary condition that $P(r = \pm 1) = 0$. Enforcing continuity of the prior at the boundary, along with the naive expectation that there should not be strong covariance between optical richness and gas mass, suggests the prior $P(r) \propto 1-r^2$.  A small correlation coefficient is expected given that the scatter in X-ray luminosity is dominated by the details of the gas state in the inner cluster core.

In Table~\ref{tab:params}, we provide a summary of our model parameters. The ``Priors" column specifies whether each parameter is taken from the \sdss\ cluster cosmology results (C19) or is drawn from a prior.  The resulting ``X-ray extended'' chains can be used to evaluate the likelihood of the clusters in our ``richest 29 clusters'' X-ray sample, thereby allowing us to use importance sampling to improve the cosmological constraints from C19.

\begin{table*}
\centering\footnotesize
\caption{$\mathcal{L}_{Chandra}$ model parameters. The parameters denoted with C19 are taken from the \sdss\ cluster cosmology chains. The gas mass scaling relation parameters were sampled from their respective prior distributions. A bracketed range indicates a uniform distribution. In addition to the parameters listed below, we recompute the halo mass function, concentration-mass relation, and survey volume using the cosmology parameters given at each link in the chain.}
\begin{tabular}{lcc}
	\hline
	Parameter & Description & Priors \\
    	\hline \vspace{-3mm}\\
	$\omm$ & The matter energy density & C19 \vspace{0.5mm}\\
	$\Omega_{\rm b}$ & The baryon energy density & C19 \vspace{0.5mm}\\
	$\Omega_{\nu}$ & The neutrino energy density & C19 \vspace{0.5mm}\\
	$\sigma_8$ & The amplitude of the matter power spectrum & C19 \vspace{0.5mm}\\
	$h$ & The Hubble parameter & C19 \vspace{0.5mm}\\
	$n_s$ & The spectral index & C19 \vspace{0.5mm}\\
	\hline
	$M_{\rm min}$ [$\massunits$] & Characteristic halo mass to form a central galaxy & C19 \vspace{0.5mm}\\
    	$M_1$ [$\massunits$] & Characteristic halo mass to acquire one satellite & C19 \vspace{0.5mm}\\
    	$\alpha_\lambda$ & Slope of richness-mass relation & C19 \vspace{0.5mm}\\
    	$\sigma^2_{\lambda,\intr}$ & Intrinsic variance in richness--mass relation & C19 \vspace{0.5mm}\\
    	$s$ & Mass function slope nuisance parameter & C19 \vspace{0.5mm}\\
    	$q$ & Mass function amplitude nuisance parameter & C19 \vspace{0.5mm}\\
	\hline
    	$A_\gas$ [$10^{15}\,M_\odot$] & Amplitude of gas mass--mass relation & $[0.5\Omega_{\rm b}/\Omega_{\rm m}, \Omega_{\rm b}/\Omega_{\rm m}]$ \\
    	$B_\gas$ & Slope of gas mass--mass relation & $[1.0, 1.4]$ \\
    	$\sigma_{\gas,\intr}$ & Intrinsic scatter in gas mass--mass relation & (0.0, 0.11] \vspace{0.5mm}\\
    	$r$ & Correlation coefficient & $\propto 1-r^2$ \vspace{0.5mm} \\
    	\hline \vspace{-3mm}\\
\end{tabular}
\label{tab:params}
\end{table*}

\section{Results}
\label{sec:results}

We used our center-corrected sample of the 29 richest \sdss\ \redmapper\ clusters, their \chandra\ X-ray gas masses, and our likelihood ($\mathcal{L}_{Chandra}$, Equation~\ref{eq:likelihood}) to importance sample the \sdss\ cluster cosmology chains from C19. For each step in the chain, we recompute the halo mass function, the concentration--mass relation, and the survey volume in the new cosmology and sample the prior distributions in Table~\ref{tab:params} for each of our X-ray parameters ($A_\gas$, $B_\gas$, $\sigma_{\gas,\intr}$, $r$). We then compute the value of our likelihood and use it as a statistical weight.

\subsection{Observable--mass constraints}
\label{sec:results:scalingrelations}
\begin{figure}
	\centering
	\includegraphics[width=\linewidth]{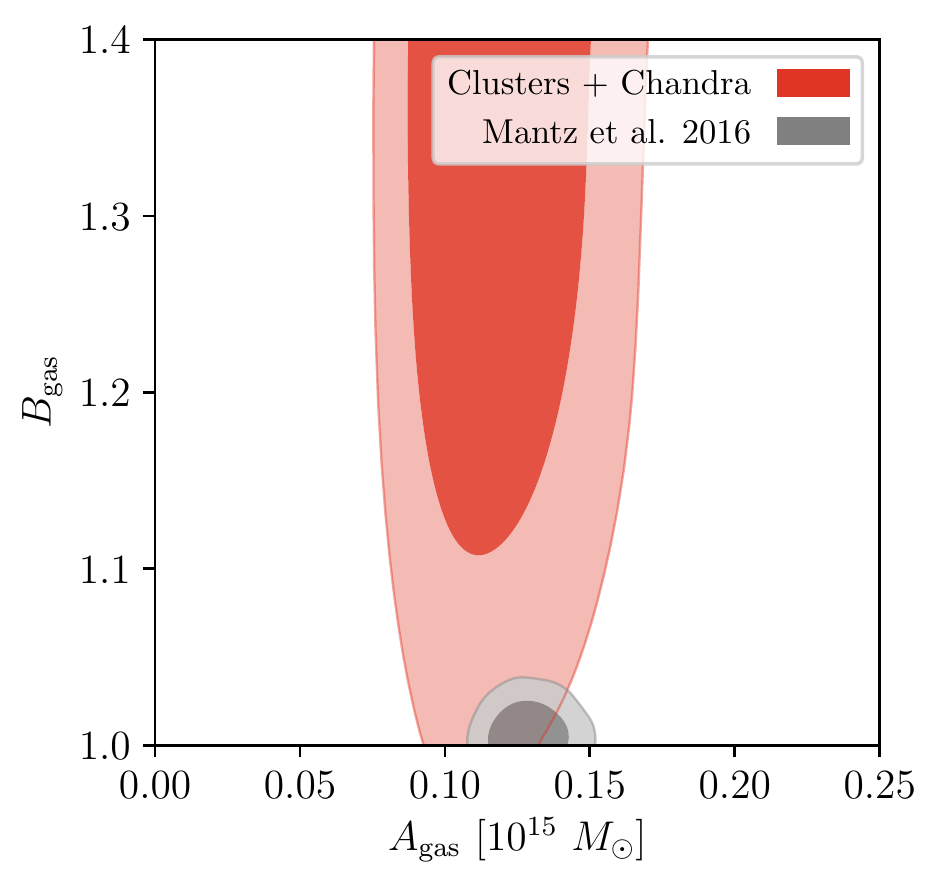}
	\caption{The $68\%$ and $95\%$ confidence contours for our posteriors on the amplitude and slope of \mgmrel. In red, we show our constraints after adopting conservative prior distributions. The gray contours show the M16 constraints for comparison. Note that while our data set does not constrain the slope of the $M_\gas$--$M$ relation, the posterior on the amplitude of the gas mass scaling relation is consistent with that from \citet{Mantz16_WtG5}, despite the two analyses relying on two very different cluster selection functions.}
	\label{fig:results:mgas}
\end{figure}

In Figure~\ref{fig:results:mgas} we show the 68\% and 95\% confidence contours for the amplitude and slope of the gas mass--mass scaling relation in red. \citet{Mantz16_WtG5}, henceforth M16, built an X-ray flux limited catalog of clusters with weak lensing measurements \citep{vonderLinden14_WtGWeakLensing, Applegate14_WtGWeakLensing} and deep \chandra\ imaging to derive X-ray observable--mass scaling relations. We show M16's constraints on the gas mass--mass relation as gray contours in Figure~\ref{fig:results:mgas} for comparison. We find that our measurement of the amplitude is in agreement with that of M16 despite the vastly different cluster selection functions between the two analyses.  This agreement should not be surprising, in that in the absence of systematics, and assuming both selection functions are adequately modeled, the results should agree.  Thus, the consistency between the two amplitude estimates is highly reassuring.  We are unable to place constraints on the slope of the $M_\gas$--$M$ relation, $B_\gas$. Future complete cluster samples that extend to lower richness will allow a direct measurement of the slope of \mgmrel, enabling a stronger consistency test. 

We also ran our analysis assuming a tight, informative prior of $B_\gas = 1.00 \pm 0.02$. This tight prior broadened our constraint on the slope of the richness--mass relation to include lower values. Since the slope of the richness--mass relation is correlated with cosmology, this prior also modestly broadened our cosmology constraints.  Overall, the addition of tight priors on the \mgmrel\ scaling relation has only a minor impact on our conclusions, so we have chosen to focus exclusively in the results derived from our conservative priors.

\begin{figure}
	\centering
	\includegraphics[width=\linewidth]{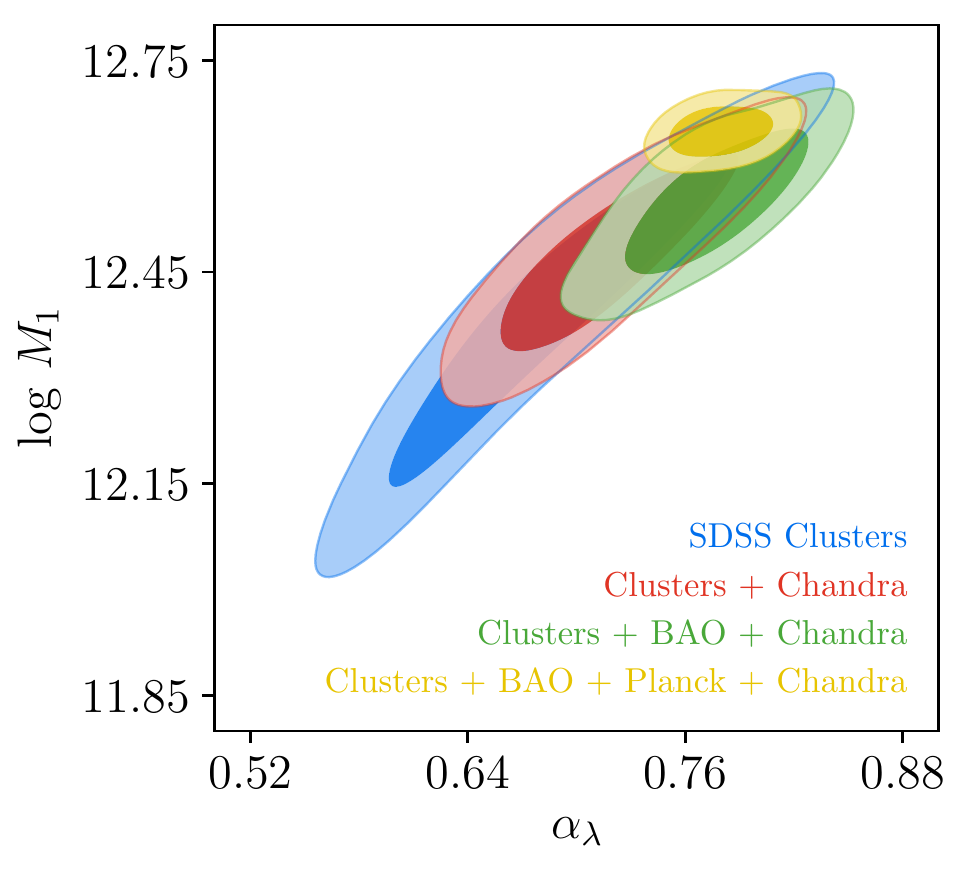}
	\caption{The 68\% and 95\% confidence contours for the well-constrained parameters of \rmrel\ before (blue) and after (red) introducing our X-ray cluster sample to the \sdss\ cluster cosmology results. We also show our constraints from the joint analysis of \sdss\ clusters + BAO + Chandra (green) and \sdss\ clusters + BAO + Planck + Chandra (gold). The remaining two parameters of the richness--mass relation, $\log M_{\rm min}$ and $\sigma_{\lambda,\intr}$ were not well-constrained and have been omitted from this figure for clarity.}
	\label{fig:results:richness}
\end{figure}

The current limiting factor governing the precision of optical cluster cosmology is the characterization of the richness--mass relation. Figure~\ref{fig:results:richness} shows the impact of our X-ray cluster sample on the posteriors for the richness--mass relation of \sdss\ \redmapper. The blue ellipses show the original \sdss\ cluster cosmology results. We show the constraints after introducing our X-ray cluster sample to the \sdss\ cluster cosmology results in red. We find that by including our X-ray sample, we tighten the 68\% confidence region of $\log M_1$ and $\alpha$ by 39.3\% and 30.8\% respectively. Our data does not strongly constrain $\sigma^2_{\lambda,\intr}$, though lower scatter values tend to be favored (Figure~\ref{fig:megacontours}). Our data does not improve upon the posterior of C19 for the parameter $\log M_{\rm min}$.

Our improvement to $\log M_1$ would appear to suggest that our small sample of X-ray clusters is providing information about the absolute mass calibration of our clusters. However, this is a side-effect of our HOD-inspired parameterization of the richness--mass relation. Since $\log M_1$ is very far from the mass pivot of our cluster sample, the parameters are correlated. This improvement in $M_1$ is due to a stronger constraint on the slope, $\alpha_\lambda$, but the mass uncertainty at the pivot point remains unchanged.

In Section~\ref{sec:results:cosmology}, we show that our cosmological constraints are consistent with those from BAO and \planck. We subsequently combine these data sets with our own to arrive at tighter posteriors for the richness--mass relation. As C19 has reported the joint constraints of the \sdss\ cluster sample with BAO and \planck\ data, we focus on the additional constraining power introduced by our complete multi-wavelength cluster sample. We find that by introducing our X-ray data to the \sdss\ clusters + BAO results (green in Figure~\ref{fig:results:richness}), we reduce the width of the 68\% confidence interval by 15.2\% and 32.7\% for $\log M_1$ and $\alpha_\lambda$ respectively, relative to the \sdss\ clusters + BAO constraints reported in C19. The inclusion of our X-ray data to the \sdss\ clusters + BAO + \planck\ analysis (yellow in Figure~\ref{fig:results:richness}) provides no additional improvement to $\alpha_\lambda$ or $\log M_1$ over C19.  We note that the inclusion of each of these external data sets to the SDSS analysis of C19 push the slope of the richness--mass relation towards unity.

\subsection{Cosmology constraints}
\label{sec:results:cosmology}

\begin{figure*}
	\centering
	\includegraphics[width=\linewidth]{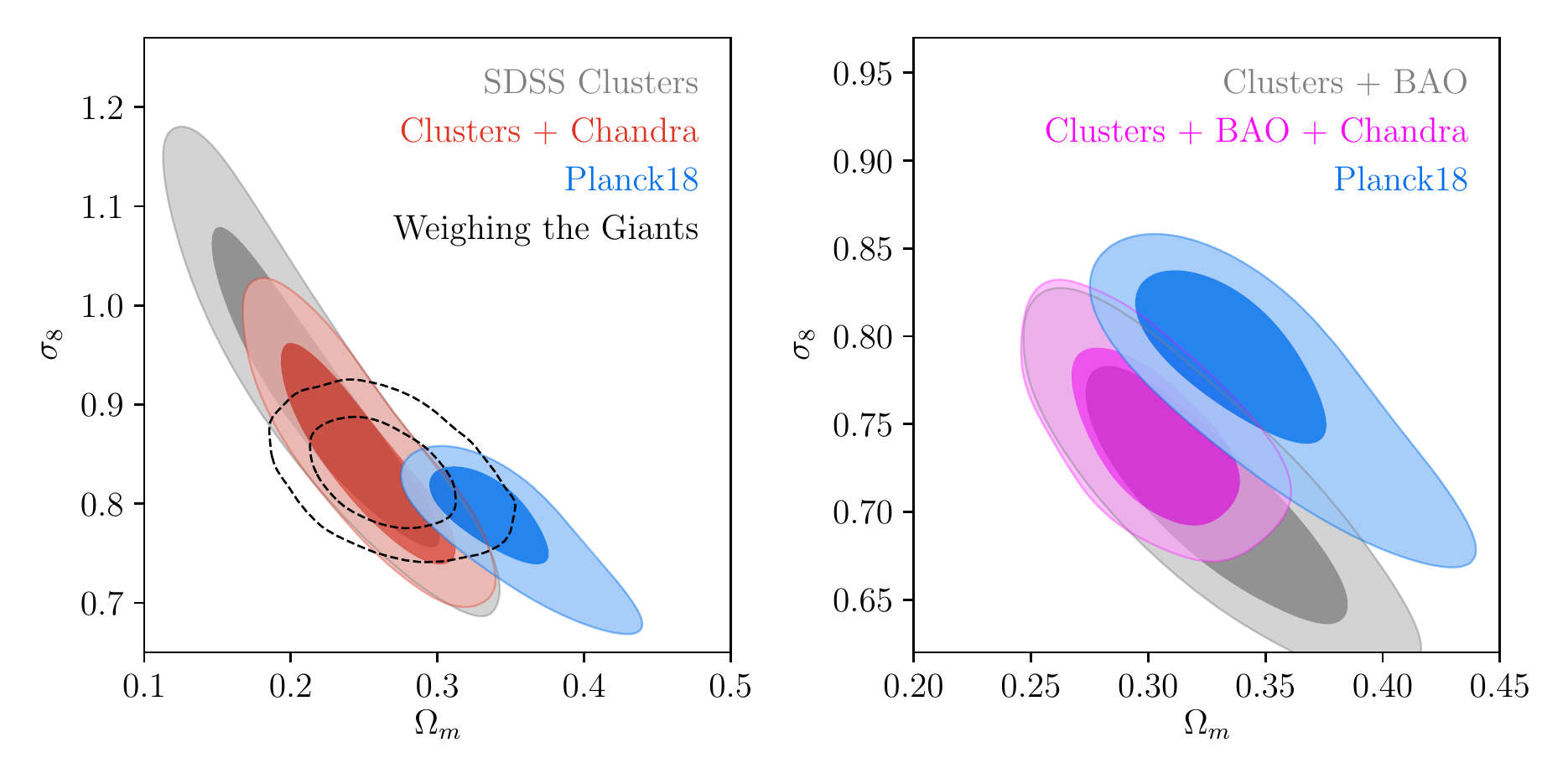}
	\caption{The impact of our X-ray cluster sample on the 68\% and 95\% confidence contours of $\sigma_8$ and $\omm$. In both panels, we show the most recent \planck\ 2018 TT+lowE results in blue for comparison. We show our constraints when we adopt our conservative priors in the red and magenta contours. \textit{Left}: The \sdss\ clusters constraints from C19 are shown in gray and the tighter measurement after introducing our X-ray clusters is shown in red. We also show the Weighing the Giants X-ray cluster cosmology constraints \citep{Mantz15_WtG4} as black dashed contours for comparison. \textit{Right}: We show the \sdss\ clusters + BAO constraints from C19 in gray and the tighter measurement after introducing our X-ray clusters in magenta.}
	\label{fig:results:sigma8-omm}
\end{figure*}

In Figure~\ref{fig:results:sigma8-omm} we show the impact of our \chandra\ data on the $\sigma_8$--$\omm$ plane. In the left panel we show the original \sdss\ cluster cosmology results in gray. In red, we show the our constraints after including our X-ray sample. We find that introducing our X-ray clusters tightens the width of the 68\% confidence interval of $\sigma_8$ and $\omm$ by 29.8\% and 25.5\% respectively, relative to the \sdss\ cluster cosmology constraints of C19.  This is an impressive improvement: reaching this level of precision through additional photometric data would have required extending the SDSS survey by an additional $\approx 5,000\ \mbox{deg}^2$. For comparison, the left panel of Figure~\ref{fig:results:sigma8-omm} shows the cosmological posteriors from \citet{Mantz15_WtG4}, which are evidently in excellent agreement with ours despite the very different selection criteria between the two analyses.

In the right panel of Figure~\ref{fig:results:sigma8-omm} we show the impact that our X-ray catalog has on the \sdss\ clusters + BAO constraints. In gray we show the \sdss\ clusters + BAO constraints from C19 and in magenta we show our improved constraints after introducing our \chandra\ data. Introducing our X-ray catalog to the \sdss\ clusters + BAO data reduces the width of the 68\% confidence interval of $\sigma_8$ and $\omm$ by 32.3\% and 36.5\% respectively, relative to the \sdss\ clusters + BAO constraints in C19.

\begin{figure*}
	\centering
	\includegraphics[width=\linewidth]{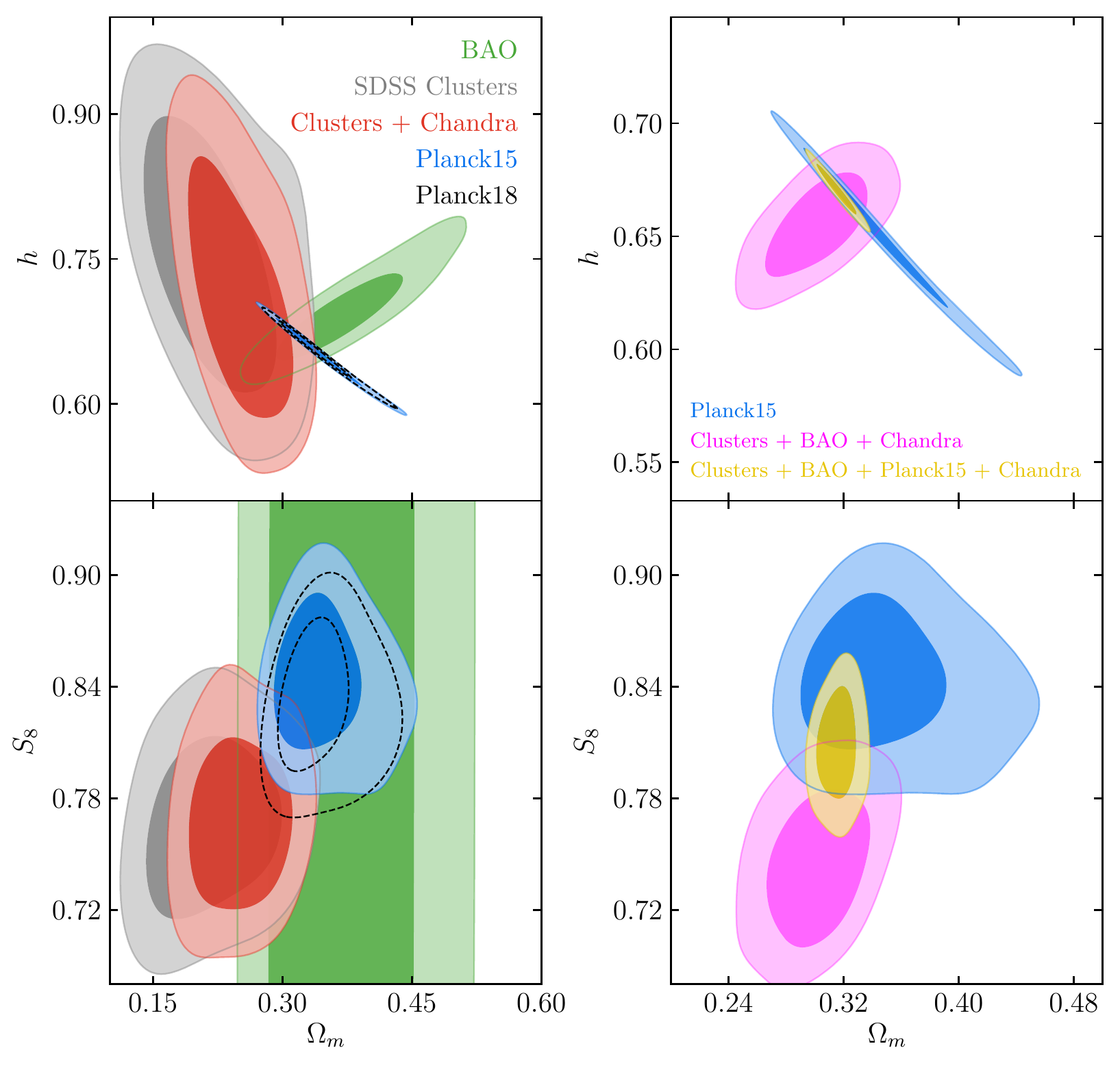}
	\caption{\textit{Left}: The impact of our X-ray cluster sample on the 68\% and 95\% confidence contours of $\omm$--$S_8$--$h$. We show the \sdss\ cluster cosmology constraints in gray. In red we show our constraints after including our \chandra\ data with the \sdss\ cluster cosmology results. We also show the constraints from our BAO data (green), \planck\ 2015 (blue), and \planck\ 2018 (black, dashed) for comparison. \textit{Right}: Our joint constraints on $\omm$--$S_8$--$h$ after including our X-ray cluster sample to the \sdss\ clusters + BAO results (magenta). We also show our joint constraints after including our X-ray cluster sample to the \sdss\ clusters + BAO + \planck\ 2015 results (gold). We show the \planck\ 2015 constraints (blue) for comparison.}
	\label{fig:results:h-S8-omm}
\end{figure*}

In Figure~\ref{fig:results:h-S8-omm}, we explore the impact that our X-ray cluster sample has on the posteriors of $\omm$--$S_8$--$h$ and show our joint constraints including CMB data from \planck. In the left panels, we compare the original \sdss\ cluster cosmology results (gray), our improved constraints after introducing our \chandra\ data (red), the constraints from just the BAO data (green), and the \planck\ 2015 and 2018 results (blue and unfilled-dashed respectively). In the right panels, we present our constraints from the joint analysis of \sdss\ clusters + BAO + \planck\ 2015 with our X-ray cluster sample. We summarize the marginalized 1 dimensional cosmology constraints in Table~\ref{tab:results:cosmology}. We find that our X-ray cluster sample does not increase the precision of the SDSS clusters + BAO + \planck\ analysis.

\begin{table*}
\centering\footnotesize
\caption{Marginalized 1D constraints before and after including our X-ray cluster sample with the SDSS cluster cosmology results \citep{Costanzi19_SDSSClusterCosmology}. We also show the \planck\ 2015 constraints \citep{Planck15_Cosmology}, the \planck\ 2018 constraints \citep{Planck18_Cosmology}, and the constraints from only our BAO data \citep{Beutler11_6dF, Ross15_SDSSGalaxyClustering, Alam17_BOSS}.}
\scriptsize
\begin{tabular}{lccccccc}
	\hline
	Data Set
		& $\Omega_m$
		& $\sigma_8$
		& $S_8$
		& $h$
		& $\log_{10} M_1\,[\massunits]$
		& $\alpha_\lambda$
		& $A_\gas\,[10^{15} \msun]$ \\
 	\hline \vspace{-3mm}\\ \planck\ 2015 
		& $0.344^{+0.017}_{-0.044}$
		& $0.790^{+0.057}_{-0.024}$
		& $0.842\pm 0.026$
		& $0.653^{+0.031}_{-0.014}$
		& \---
		& \---
		& \---
		\vspace{1.5mm} \\
	\planck\ 2018
		& $0.341^{+0.015}_{-0.035}$
		& $0.783^{+0.043}_{-0.016}$
		& $0.833\pm 0.025$
		& $0.654^{+0.025}_{-0.012}$
		& \---
		& \---
		& \---
		\vspace{1.5mm} \\
	BAO
		& $0.373\pm 0.053$
		& \---
		& \---
		& $0.694\pm0.033$
		& \---
		& \---
		& \---
		\vspace{1.5mm} \\
	\hline SDSS Clusters
		& $0.22^{+0.05}_{-0.04}$
		& $0.91^{+0.11}_{-0.10}$
		& $0.79^{+0.05}_{-0.04}$
		& \---
		& $12.42^{+0.16}_{-0.13}$
		& $0.65^{+0.05}_{-0.07}$
		& \---
		\vspace{1.5mm} \\
	Clusters + \chandra\
		& $0.25\pm 0.04$
		& $0.85^{+0.06}_{-0.08}$
		& $0.77^{+0.03}_{-0.04}$
		& \---
		& $12.48^{+0.08}_{-0.10}$
		& $0.72^{+0.04}_{-0.05}$
		& $0.118^{+0.018}_{-0.027}$
		\vspace{1.5mm} \\
	\hline Clusters + BAO
		& $0.316\pm 0.036$
		& $0.78\pm0.06$
		& $0.792^{+0.039}_{-0.037}$
		& $0.662^{+0.019}_{-0.022}$
		& $12.63 \pm 0.09$
		& $0.76 \pm 0.05$
		& \---
		\vspace{1.5mm} \\
	Clusters + BAO + \chandra\
		& $0.302\pm 0.023$
		& $0.74^{+0.03}_{-0.04}$
		& $0.745\pm 0.028$
		& $0.655^{+0.014}_{-0.016}$
		& $12.55\pm 0.07$
		& $0.78\pm 0.03$
		& $0.146^{+0.011}_{-0.013}$
		\vspace{1.5mm} \\
	\hline Clusters + BAO + \planck15
		& $0.316^{+0.010}_{-0.008}$
		& $0.81\pm 0.02$
		& $0.829^{+0.022}_{-0.020}$
		& $0.671^{+0.006}_{-0.008}$
		& $12.65 \pm 0.02$
		& $0.76 \pm 0.03$
		& \---
		\vspace{1.5mm} \\
	Clusters + BAO + \planck15 + \chandra\
		& $0.316^{+0.008}_{-0.010}$
		& $0.79\pm 0.02$
		& $0.808\pm 0.020$
		& $0.670\pm 0.007$
		& $12.65\pm 0.02$
		& $0.78\pm 0.02$
		& $0.133\pm 0.010$
		\vspace{1.5mm} \\
	\hline \vspace{-3mm}\\
\end{tabular}
\label{tab:results:cosmology}
\end{table*}


\section{Summary and Conclusions}
\label{sec:conclusions}

We constructed a likelihood for modeling joint optical and X-ray abundance data of a complete, richness-selected sample of galaxy clusters. We measured X-ray gas masses using \chandra\ for 11 of the richest clusters in \sdss\ \redmapper\ and, along with archival data, built a sample of the 30 richest clusters. Using this cluster sample, we improved the cosmological constraints presented in the \sdss\ cluster cosmology analysis of \citet{Costanzi19_SDSSClusterCosmology}.

We summarize our findings as follows:
\begin{enumerate}
\item We found agreement between our measurement of the amplitude of \mgmrel\ and the constraints presented in M16. Our measurement of the slope of \mgmrel\ is unconstrained, though it slightly favors larger values. We look towards future samples probing a broader range of richness values to enable improved measurements of the slope.

\item We presented improved constraints on the parameters of the richness--mass relation after introducing our X-ray cluster sample with conservative priors on the gas mass--mass relation. We found that the width of the 68\% confidence interval shrinks by 39.3\% and 30.8\% for $\log M_1$ and $\alpha_\lambda$ respectively, relative to the \sdss\ cluster cosmology constraints in C19.

\item We presented improved constraints on $\sigma_8$ and $\omm$ after introducing our X-ray cluster sample. Relative to the \sdss\ cluster cosmology constraints in C19, we measure widths of the 68\% confidence interval of $\sigma_8$ and $\omm$ that are 29.8\% and 25.5\% tighter respectively when adopting conservative priors on the gas mass--mass relation.

\item We presented improved constraints on $\sigma_8$ and $\omm$ obtained from our joint analysis of the SDSS clusters sample, BAO data, and our X-ray cluster sample. We found that, relative to the SDSS Clusters + BAO results from C19, including our X-ray cluster sample allowed us to tighten the 68\% confidence interval on $\sigma_8$ and $\omm$ by 32.3\% and 36.5\% respectively.
\end{enumerate}

Moving forward, our likelihood will be used as a component in the Dark Energy Survey's cluster cosmology pipeline to provide information on the scatter of the richness--mass relation. We also plan on extending our model to include observables in other wavelengths. For instance, a complete sample of \redmapper\ clusters from DES along with their tSZ signals from mm survey data such as that of the South Pole Telescope and the Atacama Cosmology Telescope will allow us to define a larger complete sample of clusters, thereby increasing the constraining power of the multi-wavelength data.   Given the large overlap between upcoming optical, X-ray, and tSZ cluster samples, the type of analysis presented here is bound to become an increasingly important component of future cluster cosmology analyses.

\section*{Acknowledgements}
\label{sec:acknowledgements}
 
We thank S. Grandis, D. Gruen, and P. Giles for providing comments and feedback on this manuscript. MK and ER were supported by the DOE grant DE-SC0015975.  ER also acknowledges funding from the Cottrell Scholar program of the Research Corporation for Science Advancement.  MK was also supported through Chandra Award Number G05-16124A. MC is supported by the ERC-StG ``ClustersXCosmo" grant agreement 716762. AvdL is supported by the US Department of Energy under award DE-SC0018053. TJ acknowledges funding from DOE grant DE-SC0010107. We acknowledge support from NASA through Chandra observing grant G05-16124.

Funding for \sdss-III has been provided by the Alfred P. Sloan Foundation, the Participating Institutions, the National Science Foundation, and the U.S. Department of Energy Office of Science. The \sdss-III web site is http://www.sdss3.org/.

\sdss-III is managed by the Astrophysical Research Consortium for the Participating Institutions of the \sdss-III Collaboration including the University of Arizona, the Brazilian Participation Group, Brookhaven National Laboratory, Carnegie Mellon University, University of Florida, the French Participation Group, the German Participation Group, Harvard University, the Instituto de Astrofisica de Canarias, the Michigan State/Notre Dame/JINA Participation Group, Johns Hopkins University, Lawrence Berkeley National Laboratory, Max Planck Institute for Astrophysics, Max Planck Institute for Extraterrestrial Physics, New Mexico State University, New York University, Ohio State University, Pennsylvania State University, University of Portsmouth, Princeton University, the Spanish Participation Group, University of Tokyo, University of Utah, Vanderbilt University, University of Virginia, University of Washington, and Yale University.

\bibliographystyle{mnras}
\bibliography{database.bib}

\appendix

\section{Derivation of the likelihood model}
\label{app:Lderivation}

We wish to determine the probability that the 30 richest clusters in the \sdss\ \redmapper\ sample have richness and $M_\gas$ values $(\lambda_1, M_{\gas,1}),\ (\lambda_2, M_{\gas,2}),\ $ etc.  We will assume the clusters have been rank-ordered by richness, so that cluster 1 is the richest cluster in the survey, cluster 2 is the $2^{nd}$ richest cluster, and so on.  

Let us consider the probability that the richest cluster has richness $\lambda_1$ and gas mass $M_{\gas,1}$.  We begin by pixelizing the observed richness and $M_\gas$ space, so that the probability of the richest cluster having observables $(\lambda_1,M_{\gas,1})$ is simply the probability that the pixel containing these values is occupied.  We assume the pixels are infinitesimal, so that every pixel contains at most one cluster.  In this limit, the probability distribution for the occupation of a pixel is a Bernoulli distribution, and therefore the probability of a pixel being occupied is equal to its expectation value.  With this framework, the probability that the richest cluster have richness $\lambda_1$ and gas mass $M_{\gas,1}$ is simply the probability that a cluster with these observables exists, times the probability that no richer cluster exists.  For the latter, the sum of random variables described by a Bernoulli distribution asymptotes to a Poisson distribution, and therefore the latter probability is given by $P(N_{\lambda > \lambda_1}= 0|\boldsymbol{p})=e^{-\langle N (\lambda > \lambda_1) \rangle}$ where the mean number count given this selection, $\langle N (\lambda > \lambda_1) \rangle$, is
\begin{equation}
    \label{eq:app:<N>_onecl}
    \langle N (\lambda > \lambda_1) \rangle = 
        \int_{\lambda_1}^\infty \dd\lambda^\obs \int_0^\infty \dd M_\gas^\obs\; 
        \frac{\dd N (\lambda^\obs, M_\gas^\obs | \boldsymbol{p})}{\dd\lambda^\obs \dd M_\gas^\obs}
\end{equation}
The integrals run over the space allowed by our selection function and we defined the differential component in equation~\ref{eq:p_obs}. We then write our likelihood given a sample that contains only the single richest cluster with richness $\lambda_1$ and gas mass $M_{\gas,1}$ as
\begin{equation}
\label{eq:app:likelihood_onecl}
    \mathcal{L}_{\rm Chandra}^{\mathrm{1\,cluster}}= 
        \left. \frac{\dd N (\lambda^\obs, M_\gas^\obs | \boldsymbol{p})}
        {\dd\lambda^\obs \dd M_\gas^\obs} \right|_{\lambda_1, M_{\gas,1}} 
        P(N_{\lambda > \lambda_1}= 0|\boldsymbol{p})
\end{equation}
where we have dropped the infinitesimal pixel size $\Delta\lambda \Delta M_\gas$ from the likelihood since these are constant.  The first term is the probability that a cluster is observed with observables $\lambda_1$ and $M_{\gas,1}$ and the second term is the probability that we measure no richer clusters.

We can readily extent this calculation to the two richest clusters in the survey, $\lambda_1$, $\lambda_2$. We begin with the likelihood for the single richest cluster, equation~\ref{eq:app:likelihood_onecl}, and repeat the process for the second richest cluster. We multiply in the probability of measuring a cluster with richness $\lambda_2$ and gas mass $M_{\gas,2}$ and the probability that we observe no clusters with richness between $\lambda_2 < \lambda < \lambda_1$. The latter term is, assuming Poisson statistics, $P(N_{\lambda_2 < \lambda < \lambda_1}= 0|\boldsymbol{p})=e^{-\langle N (\lambda_2 < \lambda < \lambda_1) \rangle}$ with an expected number count of
\begin{equation}
    \label{eq:app:<N>_twocl}
    \langle N (\lambda_2 < \lambda < \lambda_1) \rangle = 
        \int_{\lambda_2}^{\lambda_1} \dd\lambda^\obs \int_0^\infty \dd M_\gas^\obs\;
        \frac{\dd N (\lambda^\obs, M_\gas^\obs | \boldsymbol{p})}{\dd\lambda^\obs \dd M_\gas^\obs}
\end{equation}
When we write the likelihood, we take the product of these two probabilities giving us $e^{-\langle N (\lambda_2 < \lambda < \lambda_1)\rangle - \langle N (\lambda > \lambda_1) \rangle}$. Note that the only difference between $\langle N (\lambda_2 < \lambda < \lambda_1)\rangle$ and $\langle N (\lambda > \lambda_1) \rangle$ is the bounds on the $\lambda^\obs$ integrand. We rewrite this term as
\begin{align}\begin{split}
    \langle N (\lambda_2 < \lambda < \lambda_1)\rangle + \langle N (\lambda > \lambda_1) \rangle = 
    \langle N (\lambda > \lambda_2) \rangle \\
    = \int_{\lambda_2}^\infty \dd\lambda^\obs \int_0^\infty \dd M_\gas^\obs\;
    \frac{\dd N (\lambda^\obs, M_\gas^\obs | \boldsymbol{p})}{\dd\lambda^\obs \dd M_\gas^\obs}
\end{split}\end{align}
We then write the likelihood for our sample of the two richest clusters as
\begin{align}\label{eq:app:likelihood_twocl}\begin{split}
    \mathcal{L}_{\rm Chandra}^{\mathrm{2\,clusters}}= 
    & \left. \frac{\dd N (\lambda^\obs, M_\gas^\obs | \boldsymbol{p})}
        {\dd\lambda^\obs \dd M_\gas^\obs} \right|_{\lambda_1, M_{\gas,1}} 
    \left. \frac{\dd N (\lambda^\obs, M_\gas^\obs | \boldsymbol{p})}
        {\dd\lambda^\obs \dd M_\gas^\obs} \right|_{\lambda_2, M_{\gas,2}} \\
    & \times P(N_{\lambda > \lambda_2}= 0|\boldsymbol{p})
\end{split}\end{align}

From this point, it is straightforward for us to generalize this likelihood to a sample of the richest $N$ clusters, saying that the $N^{th}$ cluster in our sample has a richness $\lambda_N$.
\begin{equation}
    \label{eq:app:Lchandra}
    \mathcal{L}_{\rm Chandra} =
    e^{-\langle N(\lambda > \lambda_N) \rangle}
    \prod_{i=1}^N
    \left. \frac{\dd N (\lambda^\obs, M_\gas^\obs | \boldsymbol{p})}
        {\dd\lambda^\obs \dd M_\gas^\obs} \right|_{\lambda_i, M_{\gas,i}}
\end{equation}
We finally take the product of Equation~\ref{eq:app:Lchandra} and the likelihood in C19 to arrive at our constraints.

\section{Validation}
\label{app:validation}

We make two assumptions in our analysis: (1) That the total scatter in true richness given mass can be approximated as Gaussian with a Poisson term and intrinsic term added in quadrature. (2) That we can ignore the redshift evolution of the halo mass function and $P(\lambda^\obs|\lambda^\true, z)$ by evaluating them at the median cluster redshift. We tested that these assumptions do not bias our results by building mock cluster catalogs that do not make these assumptions, performing our analysis, and comparing our posteriors to the input model parameters.

We validate our pipeline using synthetic data generated from our model, and then verifying that the posteriors correctly reflect the underlying statistics of our mocks (e.g., that the 68\% confidence contour encompasses ``truth'' in 68\% of our mock samples). To build each mock catalog, we begin by selecting a fiducial model by drawing a set of parameters from a set of priors.  This step is critically important: if the ``truth'' is held fixed, then the priors are incorrect, and the model has no chance of reproducing the correct statistics.  Next, we build a halo catalog assuming a 10,000 sq deg survey with masses above a mass $M_\mathrm{min}$ and redshifts in $[0.1, 0.3]$ using the tinker mass function and the mass function nuisance parameters described in section~\ref{sec:lkhd:massfunc}. $M_\mathrm{min}$ is approximately the minimum mass for a halo to from a central galaxy given our \rmrel\ parameters and we take it to be 15 times smaller than the mass at which a cluster forms its first satellite galaxy. We have verified that our choice for the value of $M_{\rm min}$ has no appreciable impact on our conclusions.

We now compute the expected true satellite richness and true gas mass for each halo given its halo mass and our fiducial model. We then assign a true richness by making a Poisson draw at the expected satellite richness, add a Gaussian perturbation with $\sigma = \sigma_{\lambda,\intr} \langle\lambda^\sat|\mm, \boldsymbol{p}\rangle$, and add 1 to account for the central galaxy. We finally draw the observed richnesses using $P(\lambda_\obs|\lambda_\true, z)$ described in section~\ref{sec:lkhd:obsvar} with each halo's true richness and true redshift.  We draw the halo's true cluster gas mass from a conditional Gaussian distribution given its true richness, the fiducial correlation coefficient, and its expected true gas mass.  We draw the observed gas mass using Gaussian centered on the true gas mass with a width of 14\% of the expected gas mass. We chose 14\% to reflect the median gas mass error for our sample of the 30 richest clusters. Finally, we select the 30 richest clusters in this full catalog as our mock sample.

\begin{figure}
	\centering
	\includegraphics[width=\linewidth]{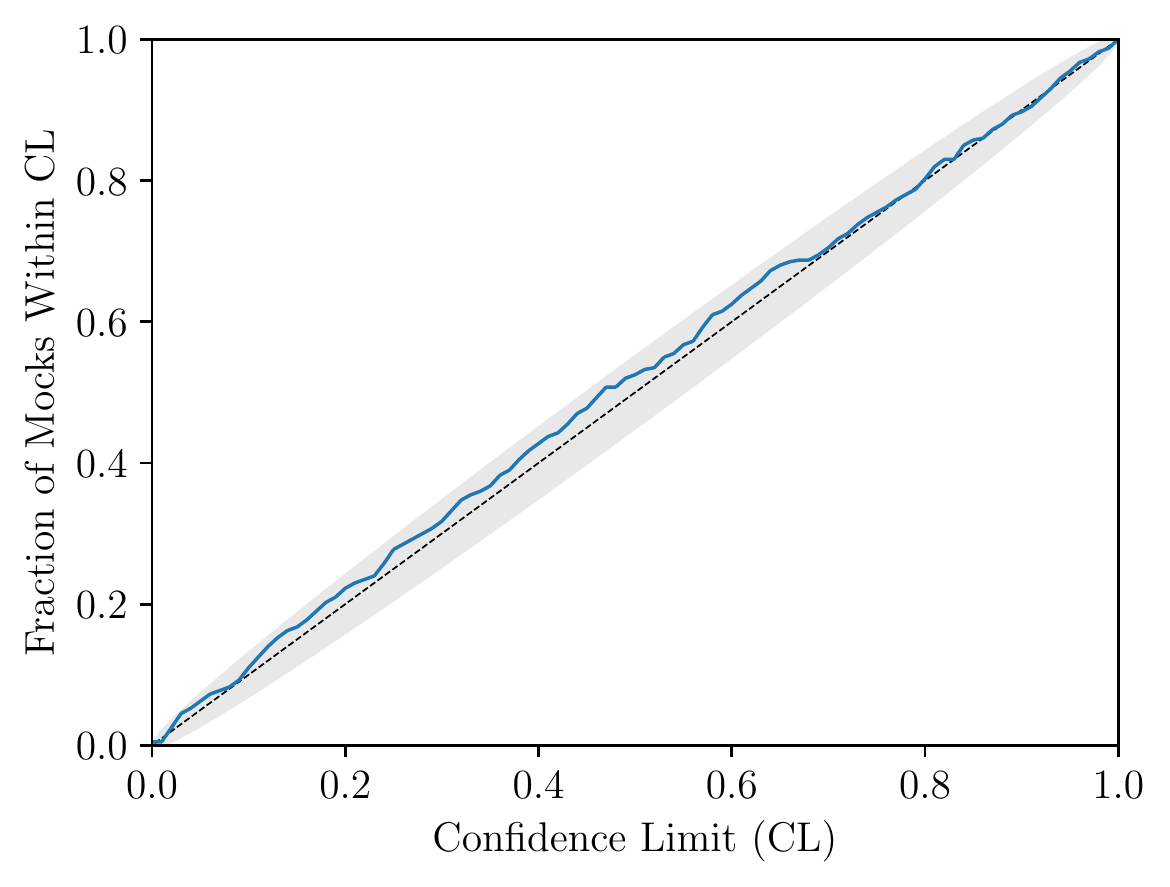}
	\caption{The fraction of mock realizations where the fiducial model falls within a given confidence limit. If our results are free of biases, the posteriors represent the true probability distributions, and we will observe the identity function. We computed the blue curve by running an MCMC with our likelihood on 500 mock catalogs. The black dashed line is the identity function and the gray shaded region is the 68\% binomial confidence region around this line.}
	\label{fig:app:coverageprob}
\end{figure}

We built an ensemble of 500 mock cluster catalogs and sampled our likelihood with each mock using the affine-invariant Markov chain Monte Carlo (MCMC) sampler \textit{emcee} \citep{emcee}. We chose an ensemble size of 500 to beat down uncertainty while remaining computationally feasible. We used 18 walkers with 2500 steps each, discarding the first 500 steps of each walker as the burn-in. This left us with posteriors that are described by Markov chains with 36,000 links which we found to be converged. Using these chains and the fiducial model for each mock, we measure the fraction of times that the truth fell within a given confidence limit. The results of this test are shown in Figure~\ref{fig:app:coverageprob}.

If our results are free of biases, we should observe the truth lying within e.g. the 68\% confidence region 68\% of the time and we should observe the identity function. We find that our results match this expectation. This allows us to conclude that our assumptions do not introduce any unintended biases to our results.

\section{Full Contour Plot}
\label{app:megacontours}
\begin{figure*}
	\centering
	\includegraphics[width=\linewidth]{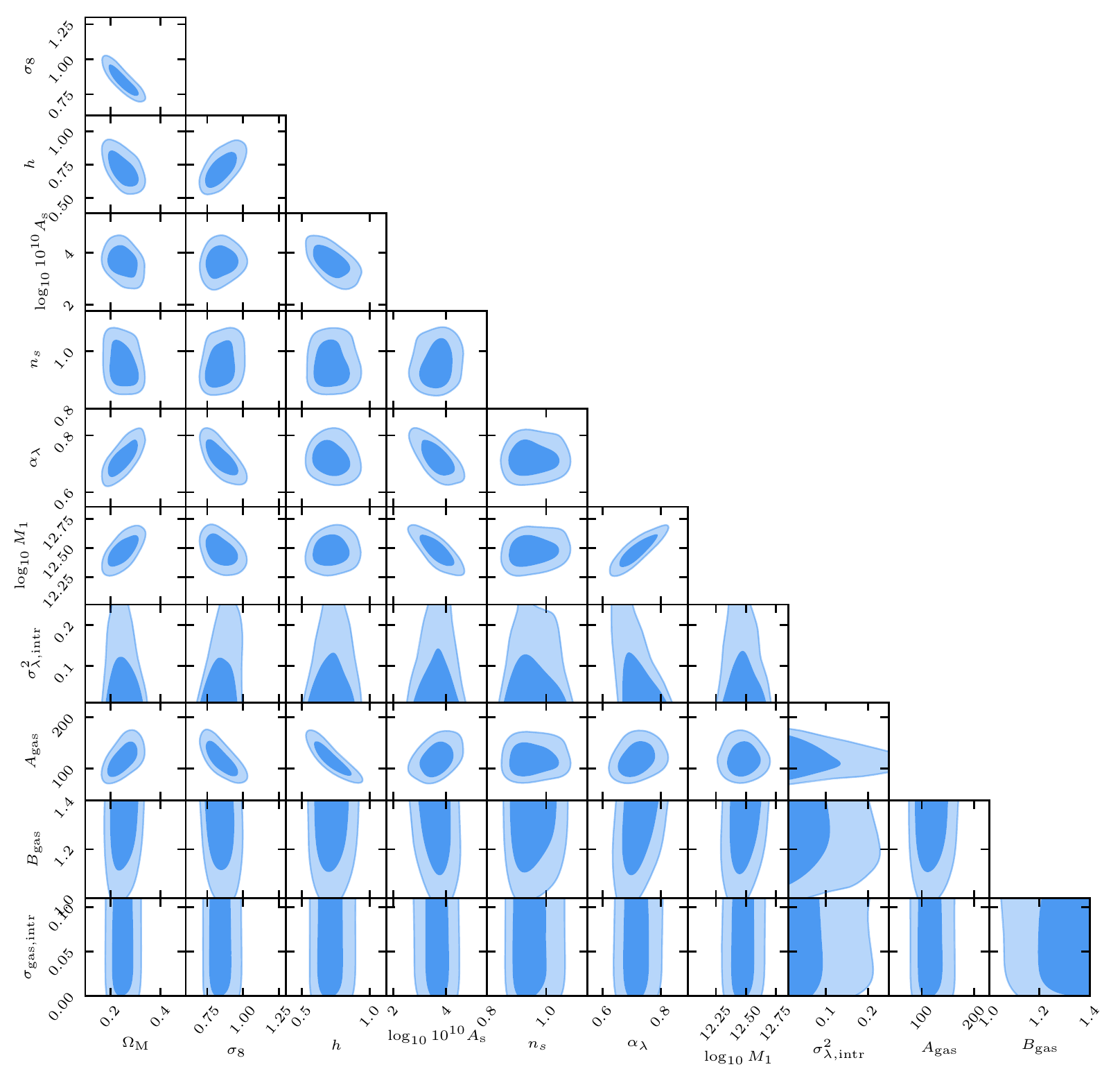}
	\caption{The 68\% and 95\% confidence contours of our posteriors from our analysis of the SDSS clusters with our X-ray cluster sample. For clarity, we have omitted three parameters that were prior dominated ($\log M_{\rm min}$, $\sigma_{\gas,\intr}$, $r$) and four parameters whose posteriors were identical to C19 ($\Omega_\nu\,h^2$, $\Omega_{\rm b}\,h^2$, $s$, $q$).}
	\label{fig:megacontours}	
\end{figure*}

In Figure~\ref{fig:megacontours}, we show a nearly complete corner plot to view correlations between parameters in our SDSS clusters + X-ray results. We have omitted three parameters that were prior dominated ($\log M_{\rm min}$, $\sigma_{\gas,\intr}$, $r$) and four parameters whose posteriors were identical to C19 ($\Omega_\nu\,h^2$, $\Omega_{\rm b}\,h^2$, $s$, $q$) for clarity.

\label{lastpage}
\end{document}